\documentclass[aps,prd,reprint,preprintnumbers,superscriptaddress,showpacs,twocolumn]{revtex4-1}
\usepackage{latexsym,graphicx,amssymb,amsmath,mathrsfs}
\usepackage{setspace,bm}
\usepackage[breaklinks, colorlinks=true, pdfstartview=FitV, linkcolor=red, citecolor=blue, urlcolor=blue]{hyperref}
\usepackage[usenames]{color}
\usepackage{latexsym}
\usepackage{epstopdf}
\usepackage{mathtools}
\usepackage{comment}
\usepackage{bbm}
\usepackage{ulem}
\usepackage{url}
\usepackage{CJKutf8}
\usepackage{xcolor}
\DeclareRobustCommand{\erase}{\bgroup\markoverwith{\textcolor{red}{\rule[.5ex]{2pt}{0.4pt}}}\ULon}

\begin{document}

\title{Path optimization method for the sign problem: Insights from random matrix models}

\author{Kouji Kashiwa}
\email[]{kashiwa@fit.ac.jp}
\affiliation{Department of Computer Science and Engineering, Faculty of Information Engineering, Fukuoka Institute of Technology, Fukuoka 811-0295, Japan}

\author{Yusuke Namekawa}
\affiliation{Department of Computer Science, Fukuyama University,
Hiroshima 729-0292, Japan}

\author{Hayato Takase}
\noaffiliation{}

\begin{abstract}
The path optimization method is applied to the Stephanov model and the chiral random matrix model, both of which share several properties with QCD, to mitigate the sign problem caused by the fermion determinant.
The Stephanov model serves as a prototypical model of finite-density QCD, while the chiral random matrix model represents an ideal system featuring the Silver Blaze phenomenon.
We show that the path optimization successfully improves the average phase factor in the Stephanov model at high chemical potential, reproducing the analytical results with reduced statistical errors.
However, it fails to improve the average phase factor in the Stephanov model at low chemical potential, as well as in the chiral random matrix model.
This tendency in the phase factor behavior seems to be closely related to the global sign problem.
\end{abstract}

\maketitle

\section{Introduction}
The Markov chain Monte Carlo (MCMC) method plays an indispensable role in investigating the fundamental properties of quantum chromodynamics (QCD).
In these MCMC calculations, expectation values are estimated using the Boltzmann weight determined by the action.
However, the action can become complex-valued under certain conditions, such as in QCD at finite chemical potential ($\mu \in \mathbb{R}$), although the partition function remains real.
This problem is well known as the sign problem; see Refs.\,\cite{deForcrand:2010ys,Nagata:2021bru,*Nagata:2021ugx} as an example.

Various methods have been proposed to control the sign problem, most notably the reweighting, Taylor expansion, and imaginary chemical potential methods.
However, these approaches are generally limited to the region $\mu/T < 1$, where $T$ denotes the temperature.
Alternatively, one can employ a contour deformation approach, which deforms the integration path of the partition function into the complexified dynamical variable plane to find a better integration path that mitigates the sign problem~\cite{Witten:2010cx}, relying on Cauchy's integral theorem.

The path optimization method is one of the contour deformation approaches, which was initially proposed in combination with the hybrid Monte-Carlo method~\cite{Duane:1987de} in Ref.\,\cite{Mori:2017pne}, and machine learning was subsequently introduced to find a better integration path than the original in Ref.\,\cite{Mori:2017nwj}.
Similar machine learning approaches have been adopted in several other studies.
For example, Ref.\,\cite{Alexandru:2017czx} employed machine learning to learn the manifold of the generalized thimble method, using the results of the latter as training data.
The details of these machine learning approaches are comprehensively reviewed in Ref.\,\cite{Alexandru:2020wrj}.

In this study, we apply the path optimization method to the Stephanov model and the chiral random matrix (ChRM) model.
The Stephanov model~\cite{Stephanov:1996ki,Halasz:1998qr} is a matrix model corresponding to QCD in the large-$N$ limit, where $N$ is the number of colors.
The complex Langevin method failed to reproduce the analytic results of this model~\cite{Bloch:2017sex}.
On the other hand, the worldvolume hybrid Monte Carlo method~\cite{Fukuma:2020fez} successfully reproduced them with reduced errors for the one-flavor case, and the contour deformation method with simple ans\"atze~\cite{Giordano:2023ppk} also worked for the two-flavor case.
The ChRM model is another matrix model with random matrix elements that shares the chiral symmetry of QCD.
This model has attracted strong attention in the context of the complex Langevin method, because the method can reproduce its analytic results at finite $\mu$ with a suitable complexification scheme~\cite{Mollgaard:2013qra,Mollgaard:2014mga,Ichihara:2016uld}.
These two matrix models provide ideal testbeds for understanding the sign problem in QCD.
We demonstrate how the sign problem in these models is mitigated by the path optimization method.
In particular, we focus on the distribution of the number density against the phase of the average phase factor to obtain deeper insights into the mechanism of the sign problem.

This paper is organized as follows.
In Sec.~\ref{sec:formulation}, we describe the formulations of both the Stephanov and ChRM models, together with an overview of the path optimization method.
Sections~\ref{sec:setup} and \ref{sec:Numerical_results} are devoted to the numerical setup and our results, respectively.
We provide a summary and outlook in Sec.~\ref{sec:summary}.

\section{Formulation}
\label{sec:formulation}

We review the formulation of the Stephanov and ChRM models with one-flavor, along with the details of the path optimization method.

\subsection{Stephanov model}

The Stephanov model treats the variables as complex matrices.
The action of the Stephanov model $S_\mathrm{S}$ at $T=0$ is written as
\begin{align}
    S_\mathrm{S} &= N \mathrm{tr} (W W^\dag) - \mathrm{Tr} \ln D_\mathrm{S},
\end{align}
where $N$ denotes an even positive integer, $W$ is the complex $N \times N$ matrix,
$D_\mathrm{S}$ is the fermion matrix constructed as
\begin{align}
D_\mathrm{S} =
\begin{pmatrix}
  m \, \mathbbm{1}_{N} & iW + C \\
  iW^\dag + C   & m \, \mathbbm{1}_{N}
\end{pmatrix},
\label{eq:matrix_S}
\end{align}
where $m$ is mass, and the matrix $C$ is given by
\begin{align}
C =
\begin{pmatrix}
  \mu \, \mathbbm{1}_{N/2} & 0 \\
  0 & \mu \, \mathbbm{1}_{N/2}
\end{pmatrix}.
\end{align}
Here, $\mathbbm{1}_{N}$ is the $N \times N$ identity matrix.
For further details on the model and its extension, see Refs.\,\cite{Stephanov:1996ki,Halasz:1998qr,Fukuma:2020fez} and Ref.\,\cite{Baranka:2024qve}, respectively.
The matrix elements of $W$ are parametrized by real variables as
\begin{align}
    W_{ab} = X_{ab} + i Y_{ab},
    \label{eq:xy}
\end{align}
with $X_{ab}, Y_{ab} \in \mathbb{R}$.
Thus, the total number of degrees of freedom of the model is $2 N^2$.

The partition function of the model is written as
\begin{align}
    {\cal Z}&= e^{N\mu^2} \int d W \, e^{-S_\mathrm{S}},
\end{align}
where $d W = dX dY$.
This model shares several properties with QCD, and the sign problem arises at finite $\mu$.
When $m=\mu=0$, the action exhibits symmetry under the chiral unitary transformation
\begin{align}
    W \to U W V^\dag,    
    \label{eq:chiral_unitary_trans}
\end{align}
where $U, V \in U(N)$; see Ref.\,\cite{Baranka:2024qve} for details.
For any values of $m$ and $\mu$, there is a symmetry in the diagonal subgroup
\begin{align}
    W \to U_\mathrm{diag} W U_\mathrm{diag}^\dag,
    \label{eq:diag_chiral_unitary_trans}
\end{align}
where $U_\mathrm{diag}$ is the diagonal unitary matrix.
With $m=\mu=0$, this symmetry is enhanced to the $U(1)$ symmetry under the transformation
\begin{align}
    W\to e^{i \phi} W,
    \label{eq:unitary_1}
\end{align}
where $\phi \in \mathbb{R}$.

The analytic result for the chiral condensate $\langle \sigma \rangle$ is given by
\begin{align}
    \langle \sigma \rangle
    &= - \frac{1}{2N} \frac{\partial \ln {\cal Z}}{\partial m}
    \nonumber\\
    &= \frac{1}{2N} \langle \mathrm{tr} D_\mathrm{S}^{-1} \rangle 
    \nonumber\\
    &=-m + \frac{\int_0^\infty d \rho\, e^{-N\rho}  I_1(2Nm\sqrt{\rho}) \sqrt{\rho} (\rho-\mu^2)^N }{\int_0^\infty d \rho\, e^{-N\rho}  I_0(2Nm\sqrt{\rho}) (\rho-\mu^2)^N},
    \label{eq:chiral_condensate}
\end{align}
and that for the number density $\langle n \rangle$ is
\begin{align}
    \langle n \rangle
    &= \frac{1}{2N} \frac{\partial \ln {\cal Z}}{\partial \mu}
    \nonumber\\
    &= \mu + \frac{1}{2N}
    \Bigl\langle \mathrm{tr} \Bigl[ D_\mathrm{S}^{-1}
    \begin{pmatrix}
        \mathbbm{1}_{N/2} & 0 \\
        0 & \mathbbm{1}_{N/2}
    \end{pmatrix}
   \Bigr] \Bigr\rangle 
    \nonumber\\
    &= \mu - \mu\frac{\int_0^\infty d \rho\, e^{-N\rho}  I_0(2Nm\sqrt{\rho}) (\rho-\mu^2)^{N-1} }
    {\int_0^\infty d \rho\, e^{-N\rho}  I_0(2Nm\sqrt{\rho}) (\rho-\mu^2)^N},
\end{align}
where $I_i$ denotes the $i$-th modified Bessel function of the first kind.
The one-dimensional integral over $\rho$ is evaluated numerically.

\subsection{Chiral random matrix model}

In addition to the Stephanov model, we consider the ChRM model.
The feature is that $\mu$ does not affect the results, such as the partition function.
The action of the ChRM model $S_\chi$ at $T=0$ is written as
\begin{align}
    S_\chi &= N \mathrm{tr} ({\cal W} {\cal W}^\dag) - \mathrm{Tr} \ln D_\chi,
\end{align}
where ${\cal W}_{ab} = \Phi_{ab} + \Psi_{ab}$, and the fermion matrix is constructed as
\begin{align}
D_\chi =
\begin{pmatrix}
  m \, \mathbbm{1}_{N} & D_{12}\\
  D_{21} & m \, \mathbbm{1}_{N}
\end{pmatrix},
\label{eq:matrix_chi}
\end{align}
where
\begin{align}
 D_{12} &= i \cosh(\mu) \Phi + \sinh(\mu) \Psi,
 \nonumber\\
 D_{21} &= i \cosh(\mu) \Phi^\dag + \sinh(\mu) \Psi^\dag,
\end{align}
with $\Phi, \Psi \in \mathbbm{C}^{N \times N}$.
The topological index is set to $0$.
The partition function of the model is written as
\begin{align}
    {\cal Z}&= \int d \Phi d\Psi \, e^{-S_\chi}.
\end{align}
For more details on the sign problem in the model, see Refs.~\cite{Mollgaard:2013qra,Mollgaard:2014mga,Ichihara:2016uld}.
The number of degrees of freedom of the model is $4 N^2$.
The non-diagonal parts are slightly different from those of the Stephanov model.
It is known that this ChRM model is equivalent to the model with
\begin{align}
 D_{12} &= \exp(\mu) \Phi - \exp(-\mu) \Psi^\dag,
 \nonumber\\
 D_{21} &= -\exp(-\mu) \Phi^\dag + \exp(\mu) \Psi,
\end{align}
where a linear transformation is used~\cite{Bloch:2012bh}.
The $\mu$-independence of the ChRM model comes from the Gaussian integral properties, where the only relevant terms have specific combinations such as $\Phi \Phi^\dag$ and $\Psi \Psi^\dag$ because the determinant term becomes
\begin{align}
    \mathrm{Tr} \ln D_\chi
    = \ln \mathrm{Det} D_\chi
    = \ln \mathrm{det} ( m^2 \mathbbm{1}_{N} - D_{12} D_{21} ).
\end{align}
Therefore, this model can be considered a simpler version of the Stephanov model at small $\mu$, where $\langle n \rangle \sim 0$, in the sense of the sign problem.

%%%%%%%%%%%%%%%%%%%%%%%%%%%%%%%%%%%%%%%%%%%%%%%%%%%%%%%%%%%%%%%%%%%%%%%%%%%%%
\begin{figure*}[t]%[H] 
 \centering
 \includegraphics[keepaspectratio, scale=0.35]{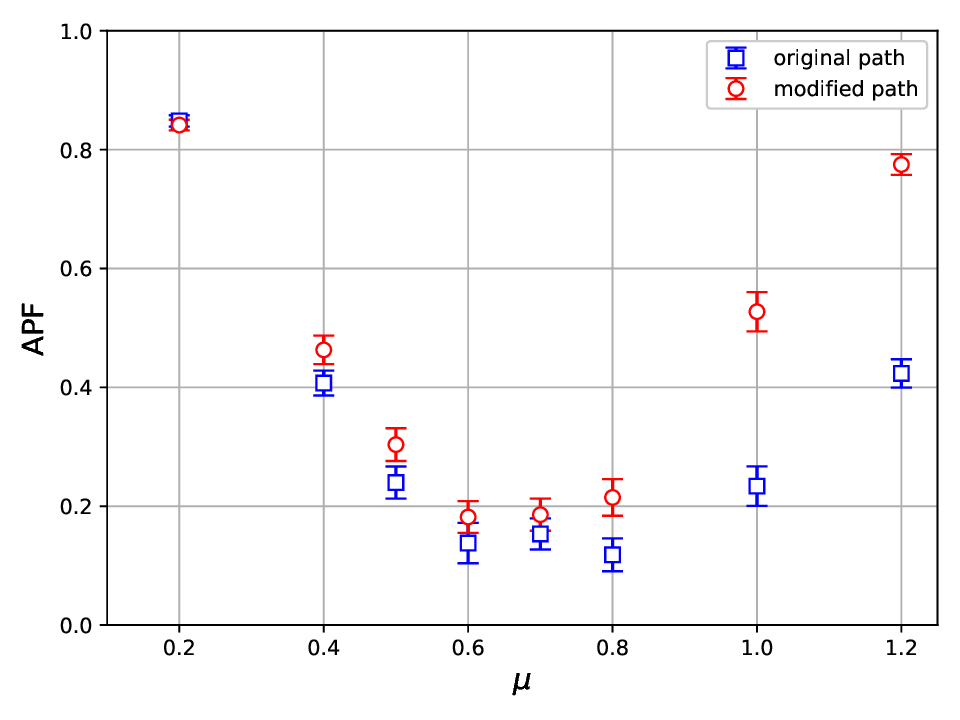}
 \includegraphics[keepaspectratio, scale=0.35]{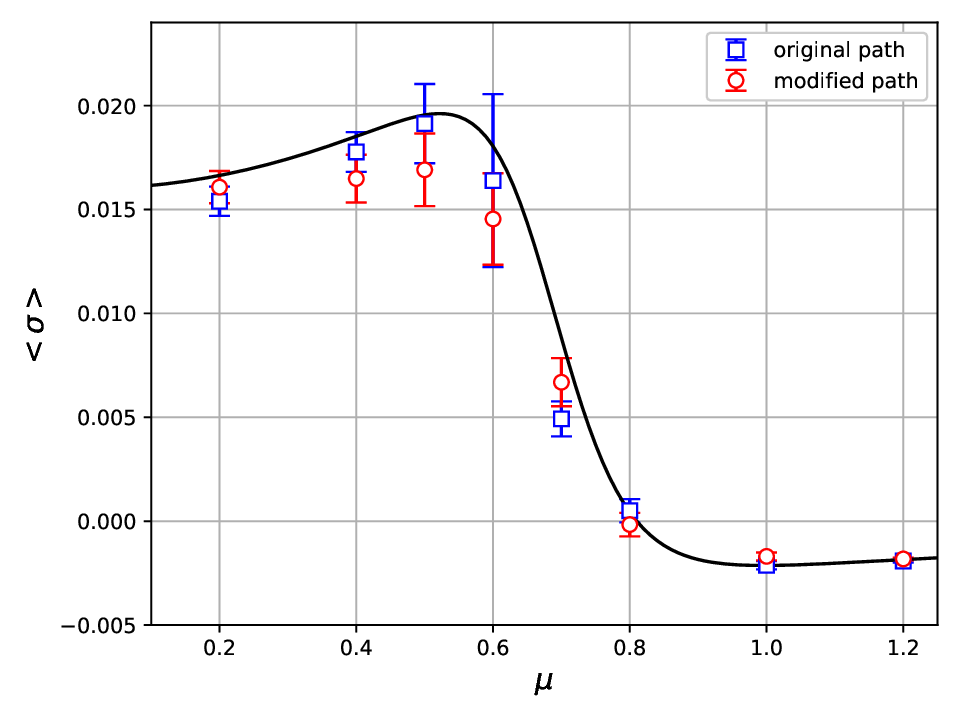}
 \includegraphics[keepaspectratio, scale=0.35]{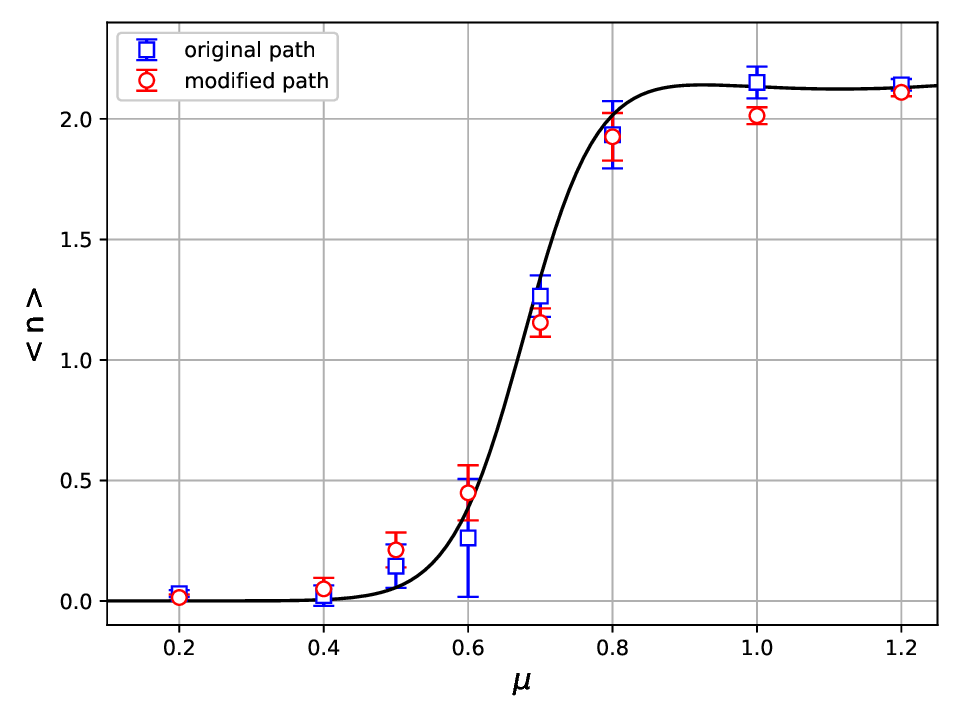}
 \caption{Expectation values as functions of $\mu$ with $N=4$ in the Stephanov model.
 The left, center and right panels show APF, $\langle \sigma \rangle$ and $\langle n \rangle$, respectively.
 The circles and squares are results on the original and modified paths, respectively.
 The solid lines are analytic results.
 }
 \label{fig:expectation_value_NN}
\end{figure*}
%%%%%%%%%%%%%%%%%%%%%%%%%%%%%%%%%%%%%%%%%%%%%%%%%%%%%%%%%%%%%%%%%%%%%%%%%%%%%

%%%%%%%%%%%%%%%%%%%%%%%%%%%%%%%%%%%%%%%%%%%%%%%%%%%%%%%%%%%%%%%%%%%%%%%%%%%%%
\begin{figure*}[t]%[H] 
 \centering
 \includegraphics[keepaspectratio, scale=0.36]{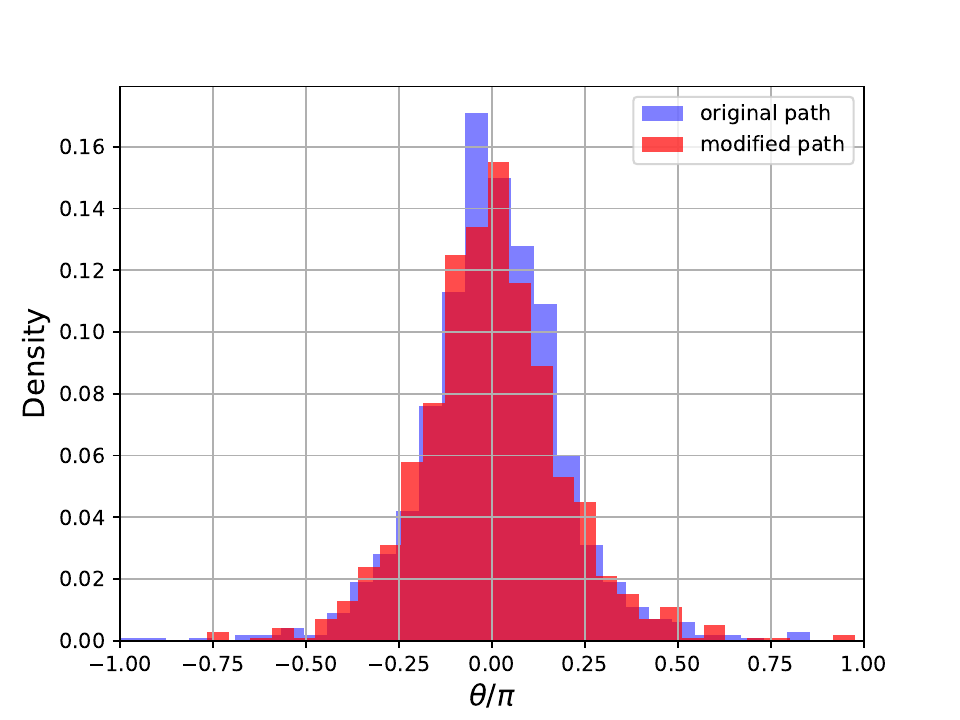}
 \includegraphics[keepaspectratio, scale=0.36]{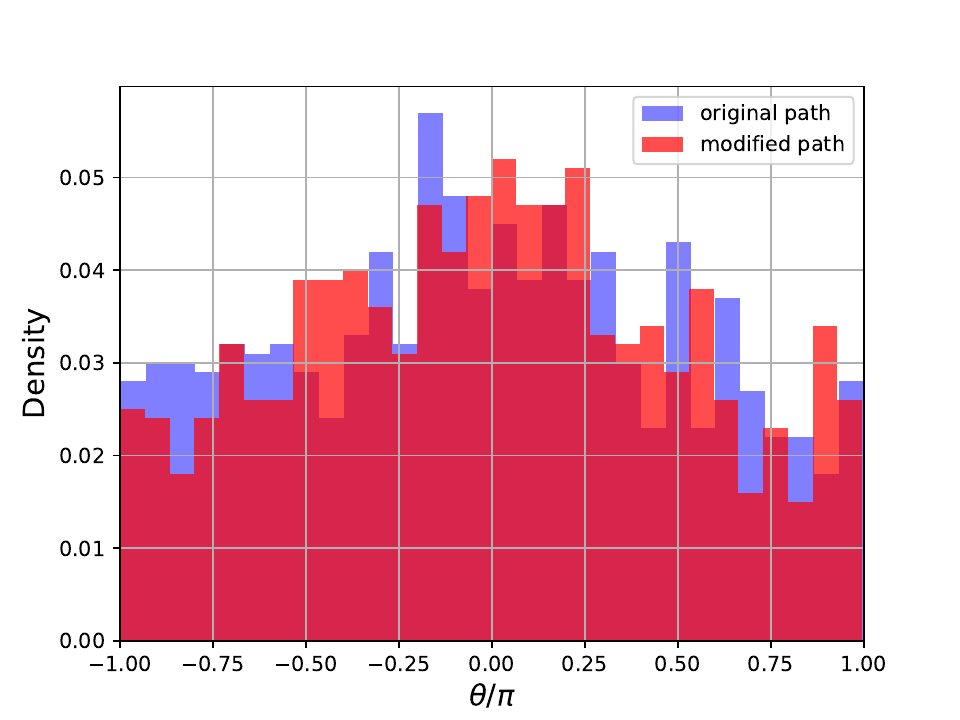}
 \includegraphics[keepaspectratio, scale=0.36]{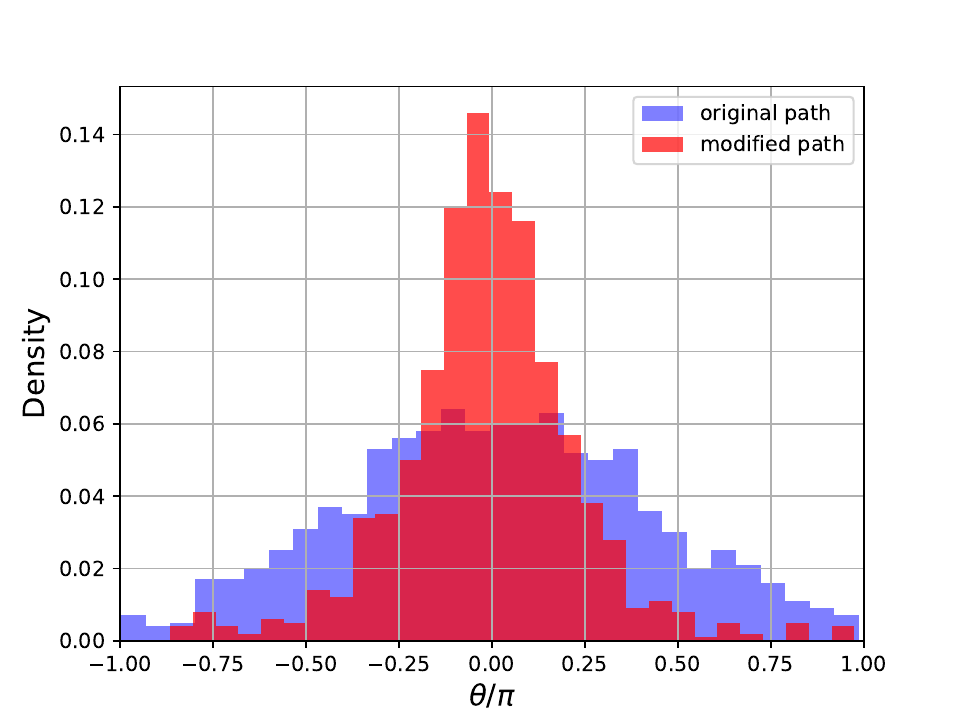}
 \caption{Normalized histogram of the phase $\theta$ on the original and modified paths with $N=4$ in the Stephanov model.
 From the left to right panels, $\mu$ is set to $\mu=0.2$, $0.6$ and $1.2$.
 }
 \label{fig:histogram}
\end{figure*}
%%%%%%%%%%%%%%%%%%%%%%%%%%%%%%%%%%%%%%%%%%%%%%%%%%%%%%%%%%%%%%%%%%%%%%%%%%%%%

%%%%%%%%%%%%%%%%%%%%%%%%%%%%%%%%%%%%%%%%%%%%%%%%%%%%%%%%%%%%%%%%%%%%%%%%%%%%%
\begin{figure*}[t]%[H] 
 \centering
 \includegraphics[keepaspectratio, scale=0.35]{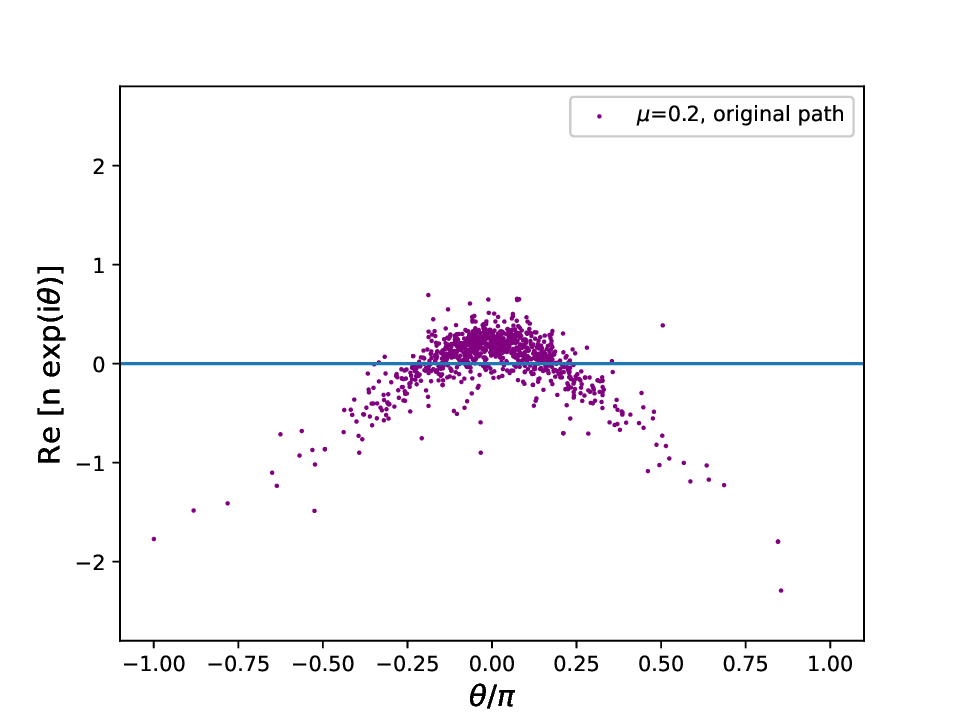}
 \includegraphics[keepaspectratio, scale=0.35]{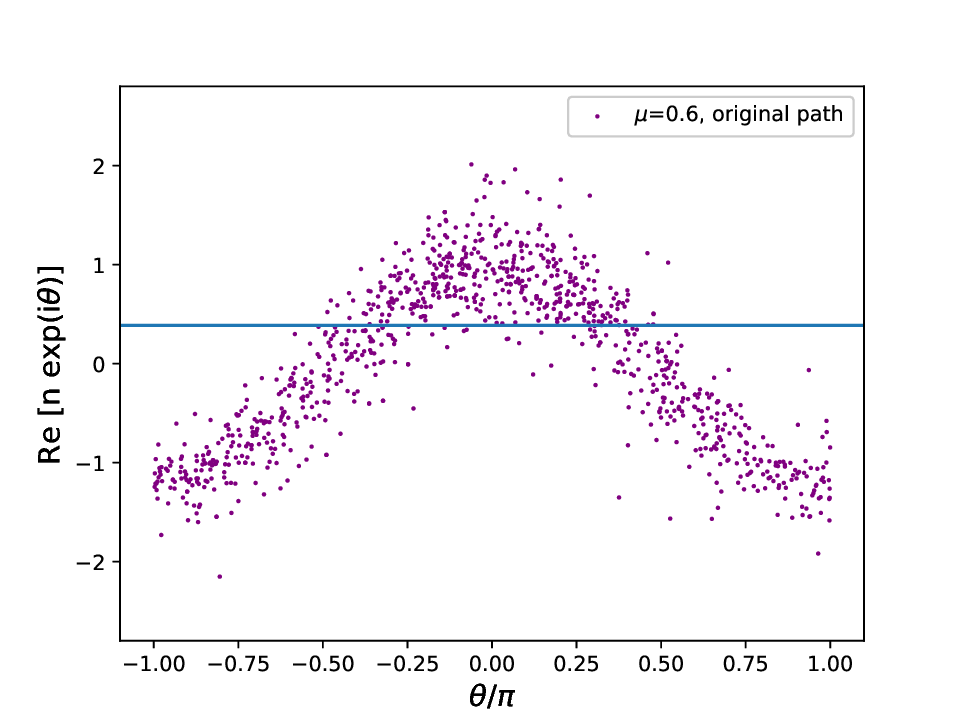}
 \includegraphics[keepaspectratio, scale=0.35]{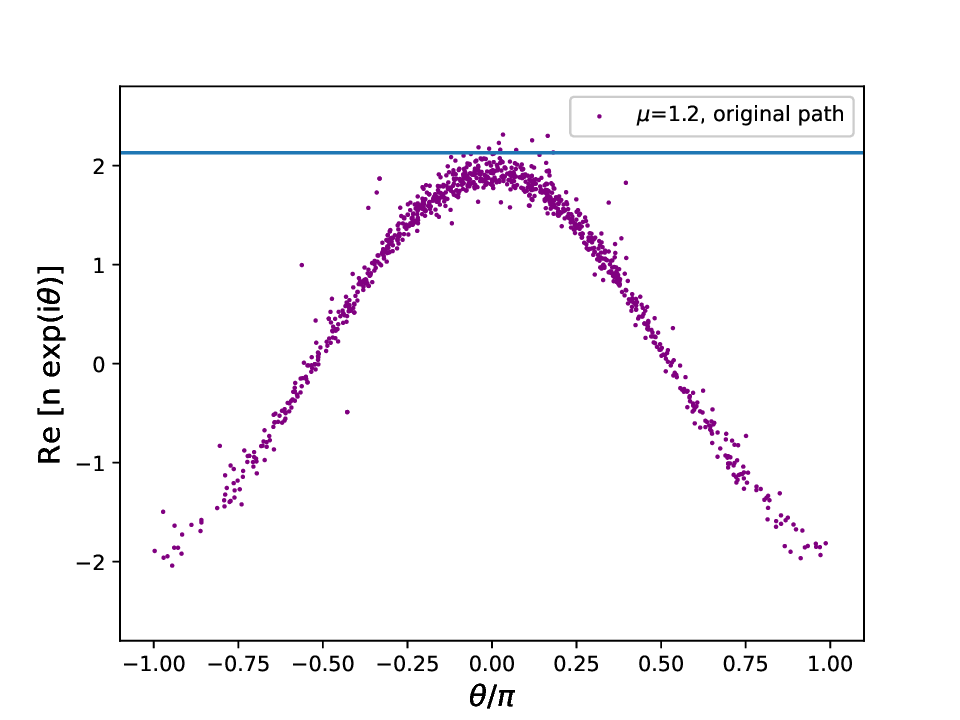}\\
 \includegraphics[keepaspectratio, scale=0.35]{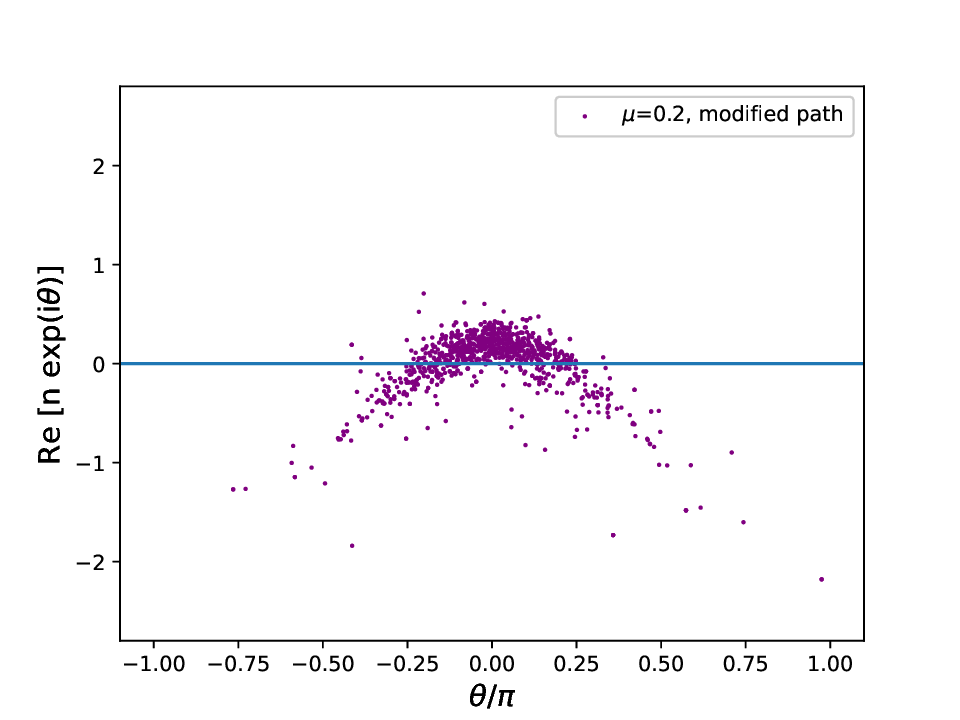}
 \includegraphics[keepaspectratio, scale=0.35]{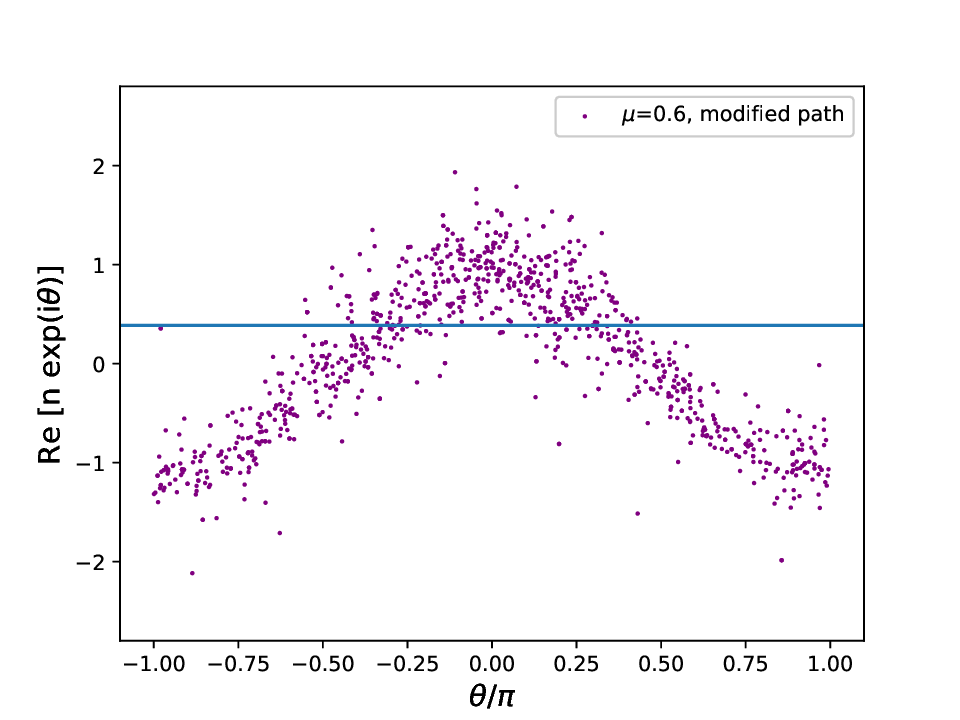}
 \includegraphics[keepaspectratio, scale=0.35]{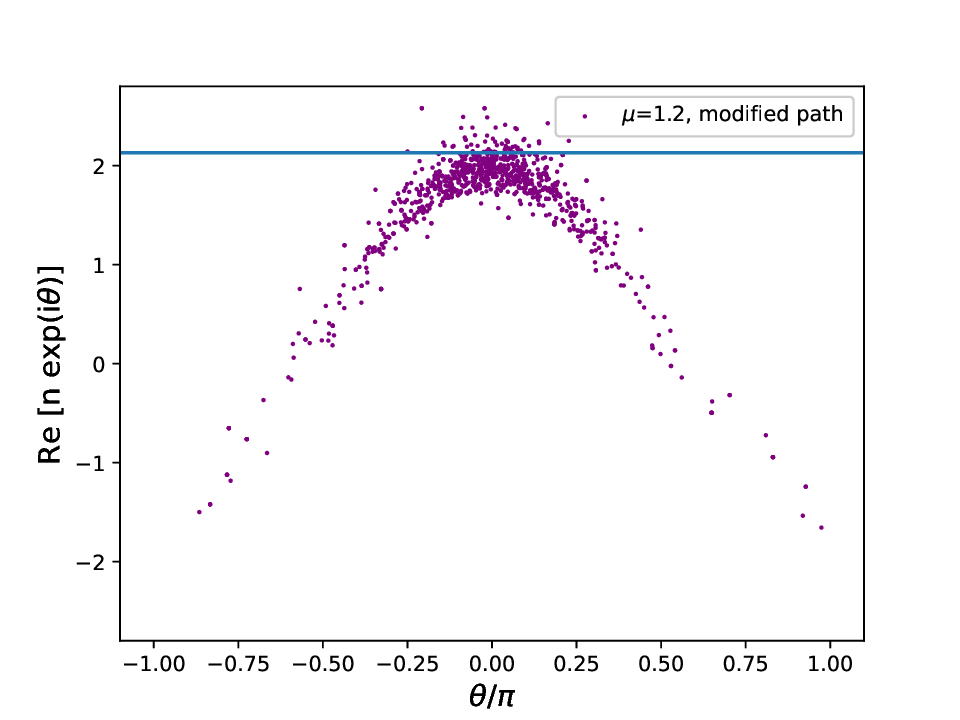}
 \caption{
 The scatter plot of $\mathrm{Re}\,[ n e^{i\theta} ]$ as a function of $\theta$ for $N = 4$ in the Stephanov model.
 The left, center and right panels show results at $\mu=0.2$, $0.6$ and $1.2$, respectively.
 The upper and lower panels show results on the original and modified paths, respectively.
 The horizontal solid lines are analytic values of $\langle n \rangle$.
}
 \label{fig:scatter_plot}
\end{figure*}
%%%%%%%%%%%%%%%%%%%%%%%%%%%%%%%%%%%%%%%%%%%%%%%%%%%%%%%%%%%%%%%%%%%%%%%%%%%%%

\subsection{Path optimization method}

Here, we briefly explain the path optimization method~\cite{Mori:2017pne,Mori:2017nwj,Alexandru:2018fqp}.
The dynamical variables $v = \{v_i\} \in \mathbb{R}^N$ are complexified as
\begin{align}
    v \to v' = v_\mathrm{R} + i v_\mathrm{I},
    \label{eq:optimized_path}
\end{align}
where $v_\mathrm{R}, v_\mathrm{I} \in \mathbb{R}^N$.
To construct a modified integration path in the complex plane, machine learning is employed.
The neural network is
\begin{align}
    \underbrace{v}_{\mathrm{input layer}} \to \mathrm{hidden layer}
      \to \underbrace{v_\mathrm{I}}_{\mathrm{output layer}}.
\label{eq:layer}
\end{align}
The feedforward operation for the hidden layers ($l = 1, \dots, L-1$) is defined as
\begin{align}
v_k^{(l)} = f \left( \sum_j w_{kj}^{(l-1)} v_j^{(l-1)} + b_k^{(l-1)} \right),
\end{align}
with the input layer set as $v_i^{(0)} = v_i$.
The output layer, which yields the imaginary part, is given by
\begin{align}
v_{\mathrm{I} l} = v^{(L)}_l = \sum_k w_{lk}^{(L-1)} v_k^{(L-1)} + b_l^{(L-1)},
\end{align}
where $w$ and $b$ denote the weights and biases of the neural network, respectively.
The function $f(\cdot)$ represents the activation function.
There are several choices for the activation function, such as the sigmoid, hyperbolic tangent, ReLU, Swish, and Mish functions; see Ref.\,\cite{szandala2020review} for a comprehensive review.
In this study, we employ the Mish function, a smooth and non-monotonic function.

To optimize the parameters of the neural network using the backpropagation method~\cite{rumelhart1986learning}, a cost function ${\cal F}$ is required.
In this work, reflecting the severity of the sign problem, we employ a cost function of the following form:
\begin{align}
    {\cal F}
    &= \frac{1}{2} \int d v_\mathrm{R} \, |e^{i\theta(v_\mathrm{R})}-e^{i \theta_0}|^2 \, |J(v_\mathrm{R}) \, e^{-S(v')}|
    \nonumber\\
     &= | {\cal Z} | \,
        \Bigl[ \langle e^{i\theta} \rangle_\mathrm{pq}^{-1} - 1 \Bigr],
    \label{eq:cost_function}
\end{align}
where $\theta = \arg(e^{-S + \ln J})$,
$\langle e^{i\theta} \rangle_\mathrm{pq}$ represents the average phase factor (APF), and $\langle \cdots \rangle_\mathrm{pq}$ denotes the phase-reweighted expectation value.
$J(v_\mathrm{R})$ is the Jacobian, and $\theta_0$ denotes the phase of the partition function.
Further details on the derivation and properties of this cost function can be found in Ref.\,\cite{Mori:2017nwj}.
Note that the explicit calculation of the Jacobian is computationally expensive; thus, reducing this computational overhead is highly desirable.
The most effective approach to mitigate this is to neglect the Jacobian during the training process.
Although this approximation slightly degrades the training quality and results in a less optimized integration path, its impact has been shown to be minor~\cite{Namekawa:2022liz,Hisayoshi:2025adi}.
Based on these findings, we adopt this Jacobian-free approximation in both the HMC update and the training process.
Finally, the expectation value of an observable ${\cal O}$ is evaluated via phase reweighting as
\begin{align}
    \langle {\cal O} \rangle
    &= \frac{\langle {\cal O} e^{i\theta} \rangle_\mathrm{pq}}
            {\langle e^{i\theta} \rangle_\mathrm{pq}}.
\label{eq:pq}
\end{align}

The flowchart of the path optimization method used in this study is illustrated in Fig.\,1 of Ref.\,\cite{Hisayoshi:2025adi}.
Although we do not employ parallel tempering~\cite{swendsen1986replica,geyer1991markov,hukushima1996exchange} here, it can be crucial for resolving the ergodicity problem in configuration sampling; for a flowchart of the path optimization method with parallel tempering, see Fig.\,1 in Ref.\,\cite{Kashiwa:2020brj}.

\section{Numerical setup}
\label{sec:setup}

%%%%%%%%%%%%%%%%%%%%%%%%%%%%%%%%%%%%%%%%%%%%%%%%%%%%%%%%%%%%%%%%%%%%%%%%%%%%%
\begin{figure*}[t]%[H] 
 \centering
 \includegraphics[keepaspectratio, scale=0.35]{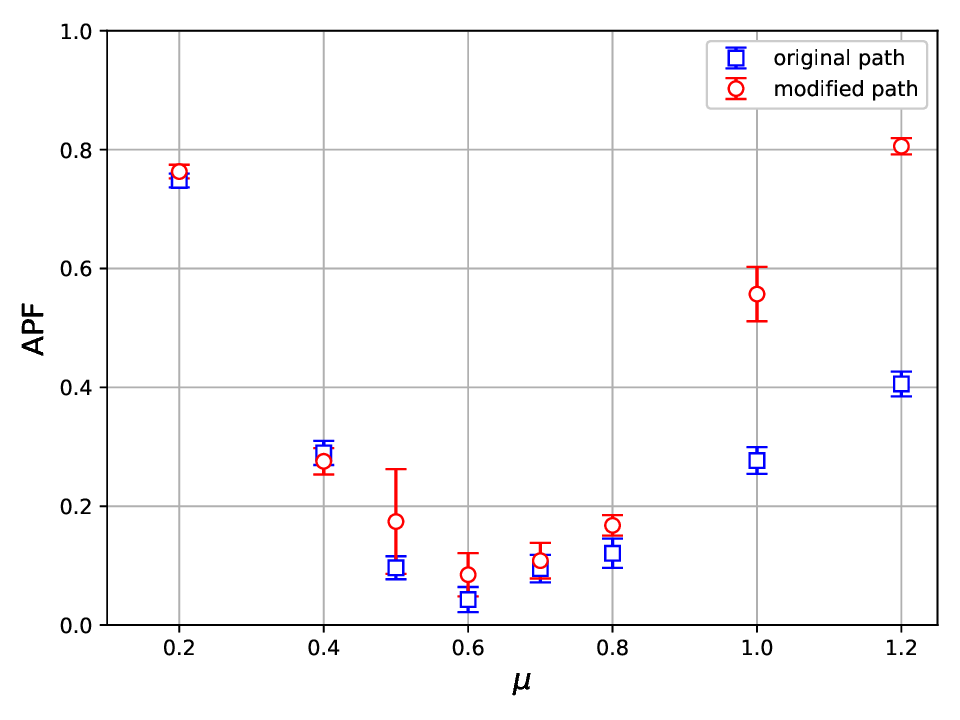}
 \includegraphics[keepaspectratio, scale=0.35]{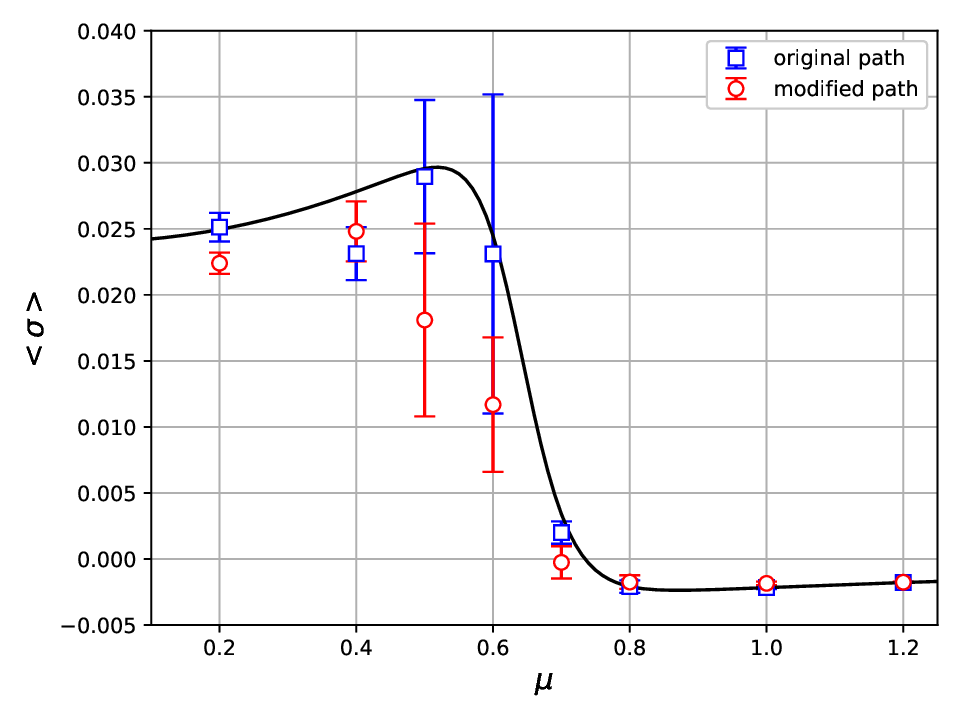}
 \includegraphics[keepaspectratio, scale=0.35]{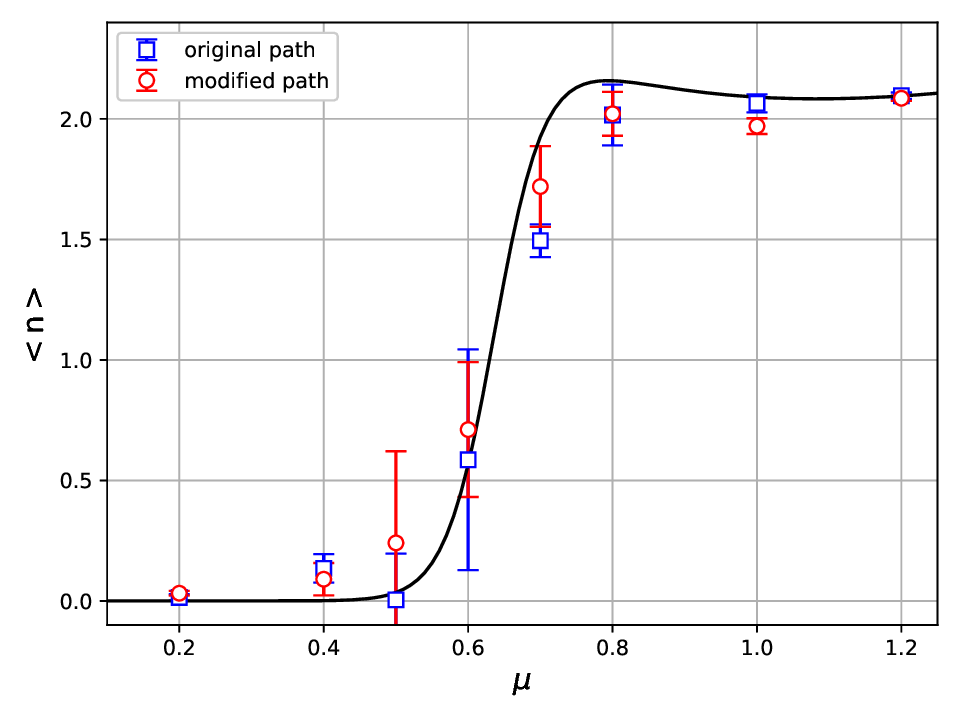}
 \caption{The expectation values as functions of $\mu$ with $N=6$ in the Stephanov model.
 The left, center and right panels show APF, $\langle \sigma \rangle$ and $\langle n \rangle$, respectively.
 The circles and squares are result on the original and modified paths, respectively.
 The solid lines are analytic results.
 }
 \label{fig:expectation_value_NN_6}
\end{figure*}
%%%%%%%%%%%%%%%%%%%%%%%%%%%%%%%%%%%%%%%%%%%%%%%%%%%%%%%%%%%%%%%%%%%%%%%%%%%%%

In this study, the mass parameter is set to $m=0.004$, which is the same value used in Ref.~\cite{Fukuma:2020fez}.
The numerical code is developed using PyTorch~\cite{paszke2019pytorch}.
We generate $1000$ configurations after thermalization using HMC within the framework of the path optimization method.

We employ a three-layer neural network with the Mish activation function~\cite{misra2019mish}.
The number of units in each layer is set equal to the degrees of freedom of the system.
We initialize the weights using the Xavier initialization~\cite{glorot2010understanding} and multiply them by $0.2$ to avoid divergence of the loss function in the early stage of training, while the biases are initialized to zero.
The parameters of the model are optimized using AdamW~\cite{loshchilov2018decoupled} and the ExponentialLR scheduler~\cite{li2019exponential} with a decay rate of $\gamma=0.9$, and the initial learning rate ranges from $10^{-4}$ to $10^{-3}$.
We perform batch training~\cite{bottou1998online} with a batch size of 64, and update the network parameters 20 times; this series of processes is referred to as "one epoch".

After training, we regenerate the configurations to address the overtraining problem.
Specifically, the training and configuration generation process is repeated for up to $20$ epochs.
Observables are then estimated using the regenerated configurations, rather than those employed for training.
The statistical error is estimated using the Jackknife method with a bin size of $50$.

\section{Numerical results}
\label{sec:Numerical_results}

In this section, we show the numerical results of the Stephanov model and the ChRM model.

\subsection{Stephanov model}
\label{sec:subsection_s}

Figure \ref{fig:expectation_value_NN} shows the APF and the expectation values of the chiral condensate and the number density with $N=4$ as functions of $\mu$.
On the original path, the APF in the range of $\mu = 0.6 - 0.8$ is small, leading to large statistical errors in the observables.
On the modified path, the APF is enhanced for all $\mu$ except $\mu = 0.2$, where the sign problem is already weak on the original path.
This enhancement increases as $\mu$ becomes large, which is consistent with the tendency reported in Refs.~\cite{Fukuma:2020fez,Giordano:2023ppk}.
Consequently, it leads to a reduction of statistical errors in the observables.

Figure~\ref{fig:histogram} shows histograms of $\theta$ on the original and modified paths with $N = 4$ at $\mu = 0.2, 0.6$ and $1.2$, respectively.
The results demonstrate that the path optimization improves the localization of $\theta$, and the improvement is particularly pronounced at large $\mu$.
Note that the imaginary part of the APF is zero because the partition function of this model is real. This can also be seen from the tendency of the histograms to become symmetric around $\theta=0$. This also holds true for the ChRM model, and thus we discuss only the real part of the APF in the following.

To understand the improvement of the APF, we present scatter plots of the number density as a function of $\theta$ in Fig.\,\ref{fig:scatter_plot}, where each data point is calculated from the generated configurations.
Data at $\mu = 0.2$ are widely scattered on both the original and modified paths, and therefore no clear improvement is achieved.
The analytic value of the number density is realized only through a severe cancellation of the positive and negative values of the data.
The situation is similar at $\mu = 0.6$.
This cancellation is regarded as a type of Silver Blaze phenomenon~\cite{Cohen:2003kd} induced by the global sign problem.
In contrast, while the data are spread widely on the original path at $\mu = 1.2$, those on the modified path are gathered around $\theta = 0$ and are close to the analytic value.
When the cancellation of the positive and negative contributions as described above is necessary to reproduce the analytic value, simply attempting to gather configurations at $\theta=0$ by modifying the integration path is ineffective; this is likely why the path optimization method under the present setup cannot improve the APF at low $\mu$.
Such a lack of improvement at low $\mu$ was also observed in Refs.\,\cite{Fukuma:2020fez,Giordano:2023ppk}, and thus this tendency is not due to a lack of expressive power of the present neural network, but is related to the Silver Blaze phenomenon.

In addition to $N=4$, we perform a similar analysis with $N = 6$ to investigate the $N$-dependence.
In Fig.\,\ref{fig:expectation_value_NN_6}, numerical results for the APF, $\langle \sigma \rangle$ and $\langle n \rangle$ are plotted, respectively.
The improvement of the APF is less clear, except at $\mu = 1.0$ and $1.2$, due to a more serious sign problem.
In addition, the ergodicity problem becomes serious, and the results occasionally deviate from the analytical values.
For better control of the sign and ergodicity problems, particularly for the region where the Silver Blaze phenomenon occurs, we may need to extend the cost function or modify the action itself; see Refs.\,\cite{Doi:2017gmk,Fukuma:2017fjq,Lawrence:2022dba}.
We will discuss these issues elsewhere.

\subsection{Chiral random matrix model}

The ChRM model has the sign problem at finite values of $\mu$ as in the case of the Stephanov model.
However, the two models differ in the $\mu$-dependence of their expectation values.
In particular, the expectation values in the ChRM model do not depend on $\mu$, and thus the expectation value of the number density remains zero.
Consequently, the ChRM model is useful to investigate how the data behave when $\langle n \rangle = 0$ in the presence of the sign problem.

\begin{figure}[t]%[H] 
 \centering
 \includegraphics[keepaspectratio, scale=0.25]{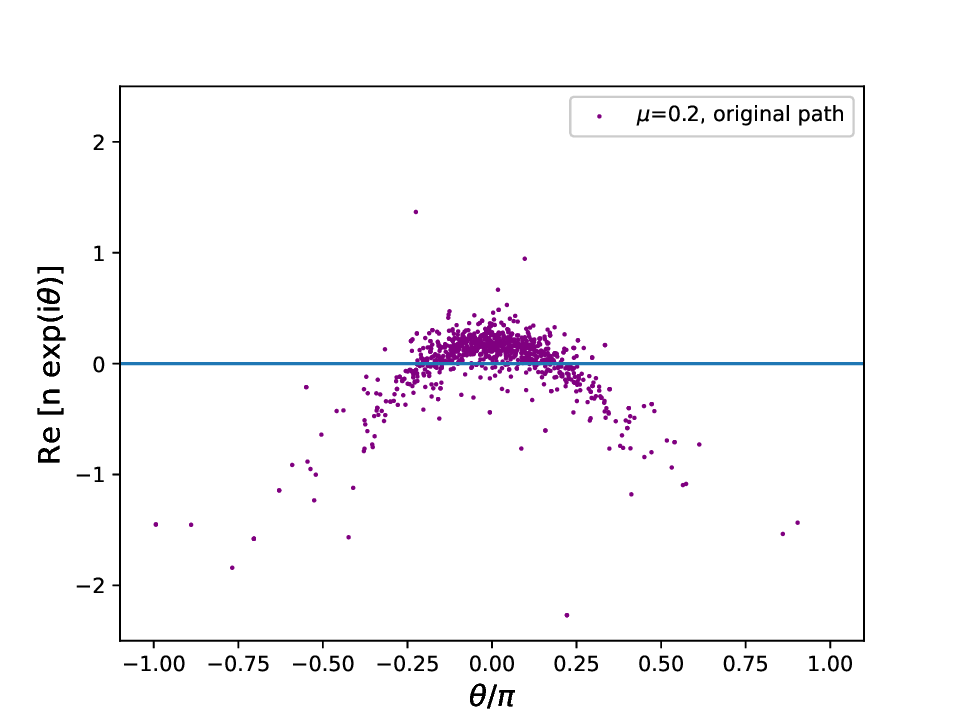}
 \includegraphics[keepaspectratio, scale=0.25]{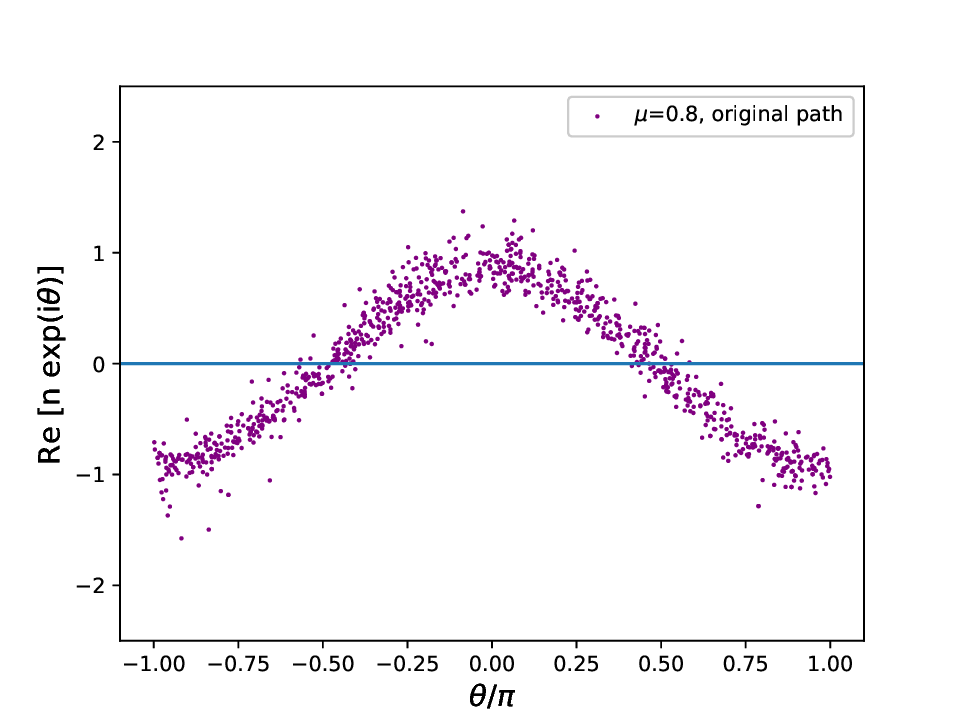}\\
 \includegraphics[keepaspectratio, scale=0.25]{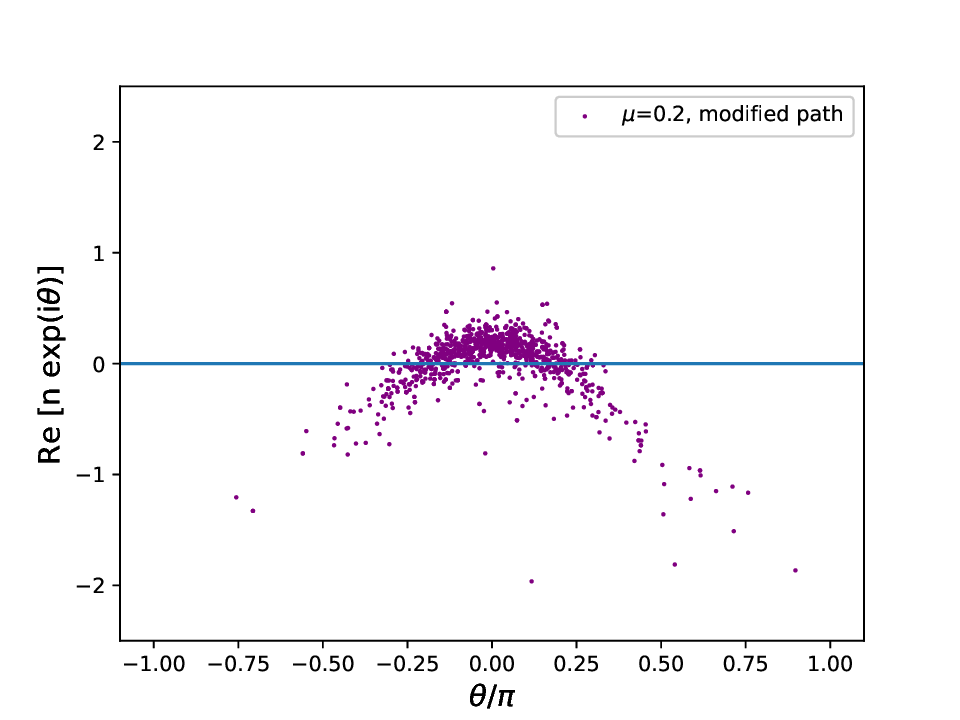}
 \includegraphics[keepaspectratio, scale=0.25]{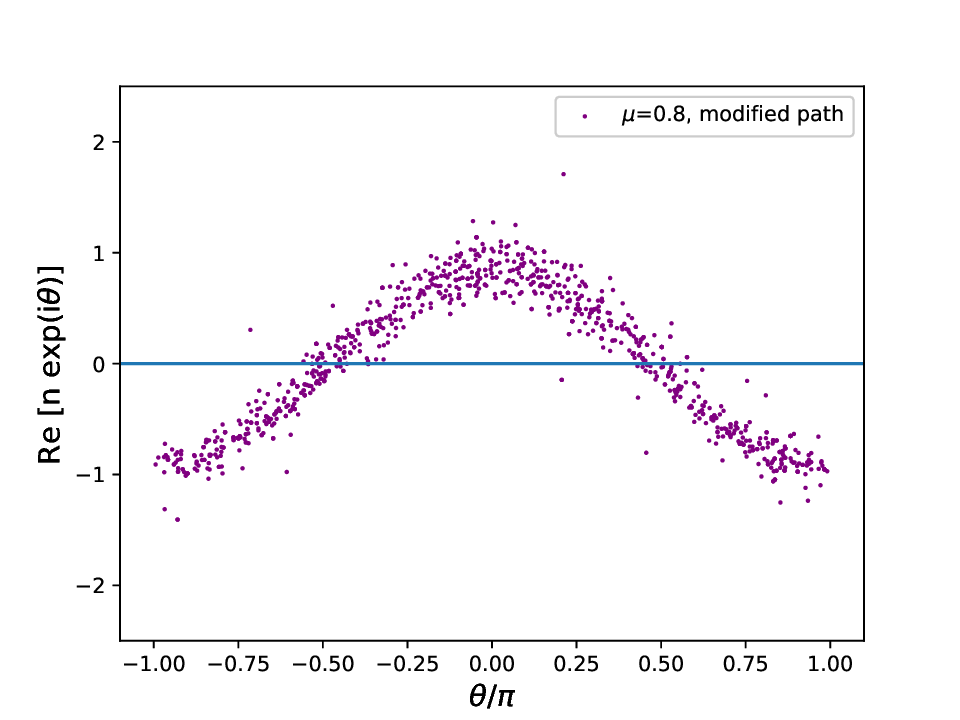}
 \caption{
 The scatter plot of $\mathrm{Re}\,[n e^{i\theta} ]$ as a function of $\theta$ for $N = 4$ in the ChRM model.
 The left and right panels show results at $\mu=0.2$ and $0.8$, respectively.
 The upper and lower panels show results on the original and modified paths, respectively.
 The horizontal lines are analytic values of $\langle n \rangle$.
}
 \label{fig:ChRM_scatter_plot}
\end{figure}

The scatter plots of the number density as a function of $\theta$ with $N = 4$ are shown in Fig.\,\ref{fig:ChRM_scatter_plot}.
The results are similar to those of the Stephanov model.
Specifically, a severe cancellation between the positive and negative contributions is required to yield $\langle n \rangle = 0$.
As in the case of the Stephanov model at low $\mu$, we do not observe a clear improvement in the APF of the ChRM model at any $\mu$,  because the same cancellation mechanism occurs.

%%%%%%%%%%%%%%%%%%%%%%%%%%%%%%%%%%%%%%%%%%%%%%%%%%%%%%%%%%%%%%%%%%%%%%%%%%%%%
%%%%%%%%%%%%%%%%%%%%%%%%%%%%%%%%%%%%%%%%%%%%%%%%%%%%%%%%%%%%%%%%%%%%%%%%%%%%%

%%%%%%%%%%%%%%%%%%%%%%%%%%%%%%%%%%%%%%%%%%%%%%%%%%%%%%%%%%%%%%%%%%%%%%%%%%%%%
\begin{figure}[t]%[H] 
 \centering
 \includegraphics[keepaspectratio, scale=0.5]{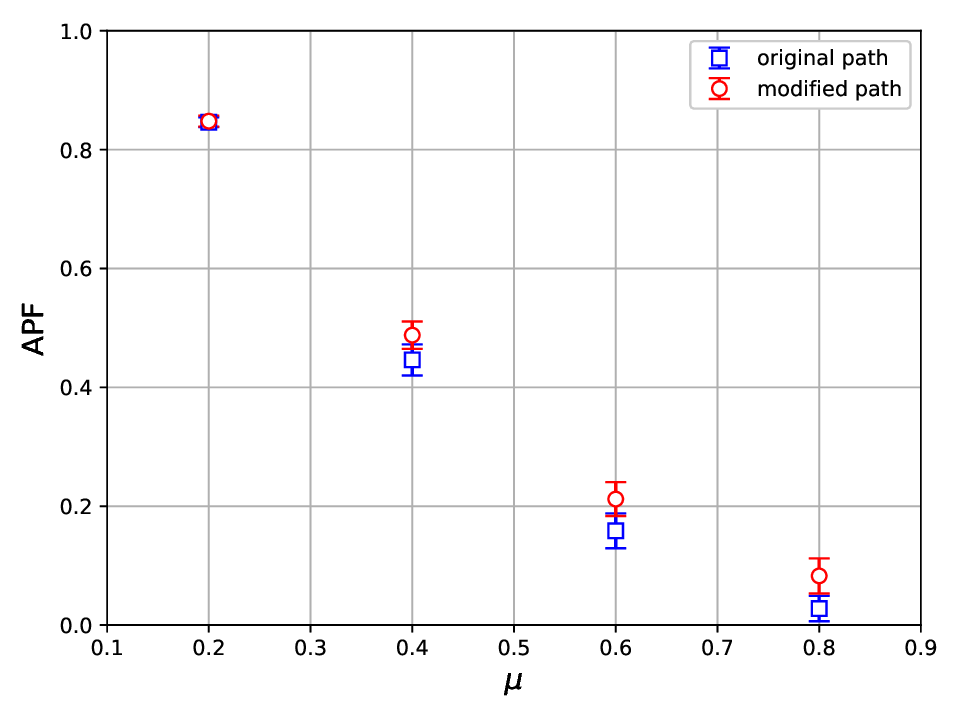}
 \caption{
 The $\mu$-dependence of APF with $N=4$ in the ChRM model.
 The circles and squares are result on the original and modified paths, respectively.
 }
 \label{fig:ChRM_APF_N}
\end{figure}
%%%%%%%%%%%%%%%%%%%%%%%%%%%%%%%%%%%%%%%%%%%%%%%%%%%%%%%%%%%%%%%%%%%%%%%%%%%%%

Figure \ref{fig:ChRM_APF_N} shows the $\mu$-dependence of APF with $N = 4$.
The sign problem becomes more severe as $\mu$ increases.
The path optimization does not lead to significant enhancement of the APF.
Therefore, as in the case of the Stephanov model, further methodological improvements are required.

\section{Summary}
\label{sec:summary}

In this study, we applied the path optimization method to the Stephanov and the chiral random matrix (ChRM) models at zero temperature and finite chemical potential ($\mu$) as a testbed to investigate the sign problem that arises from the fermion determinant term.
The integration path for the complexified variables is constructed using a neural network via self-supervised learning.

We found that the average phase factor was enhanced on the modified integration path constructed by the path optimization method at high $\mu$ in the Stephanov model, and that the numerical results for the chiral condensate and the number density reproduced the analytic results with small statistical errors.
On the other hand, at moderate and small $\mu$, the enhancement of the average phase factor on the modified integration path was less prominent, and the statistical errors were not significantly reduced.

The scatter plots of the number density as a function of the phase of the average phase factor ($\theta$) show that the contributions from positive and negative values cancel each other out to yield a small value of the expectation value of the number density at low $\mu$.
Therefore, simply attempting to gather configurations at $\theta=0$ via the modification of the integration path is ineffective.
This is likely the reason why the path optimization method cannot significantly improve the average phase factor at low $\mu$.
This situation is similar to the case of the ChRM model in which the contributions of $\mu$ to the partition function vanish, and thus the number density becomes $0$ at any $\mu$.
We may need further improvements to the path optimization method, such as extending the cost function or modifying the action, especially in the region where the global sign problem is severe.
These issues will be addressed in future studies.

\begin{acknowledgments}
This work is supported by the Japan Society for the Promotion of Science (JSPS) KAKENHI Grant Numbers (JP22H05112 and JP24K07052). 
\end{acknowledgments}

\bibliography{ref.bib}

\end{document}